# Team size matters: Collaboration and scientific impact since 1900

Vincent Larivière [1,2], Cassidy R. Sugimoto [3], Andrew Tsou [3], and Yves Gingras [2]

1. École de bibliothéconomie et des sciences de l'information, Université de Montréal, C.P. 6128, Succ. Centre-Ville, Montréal, QC. H3C 3J7, Canada

2. Observatoire des Sciences et des Technologies (OST), Centre Interuniversitaire de Recherche sur la Science et la Technologie (CIRST), Université du Québec à Montréal, CP 8888, Succ. Centre-Ville, Montréal, QC. H3C 3P8, Canada

3. School of Informatics and Computing, Indiana University, 1320 E. 10thSt, Bloomington, IN. 47405, USA

[vincent.lariviere@umontreal.ca; sugimoto@indiana.edu; atsou@indiana.edu; gingras.yves@uqam.ca]

**Abstract**
This paper provides the first historical analysis of the relationship between collaboration and scientific impact, using three indicators of collaboration (number of authors, number of addresses, and number of countries) and including articles published between 1900 and 2011. The results demonstrate that an increase in the number of authors leads to an increase in impact–-from the beginning of the last century onwards—and that this is not simply due to self-citations. A similar trend is also observed for the number of addresses and number of countries represented in the byline of an article. However, the constant inflation of collaboration since 1900 has resulted in diminishing citation returns: larger and more diverse (in terms of institutional and country affiliation) teams are necessary to realize higher impact. The paper concludes with a discussion of the potential causes of the impact gain in citations of collaborative papers.

# Introduction

The notion of the lone genius is one of science's keystone myths, simultaneously romantic and tidy. The quintessential example is that of Einstein conducting cutting-edge research while working as an examiner at the Bern patent office (Pyenson, 1985; Simonton, 2013). However, as with many myths, the "lone genius" legend is not entirely accurate. Scientific research has never been a strictly individual enterprise (Shapin, 1989), and even Einstein collaborated on several papers (Pyenson, 1985). In chemistry for example, one third of published papers had more than one author in 1900, a proportion that grew to 70% by the end of World War II (Gingras, 2010). In contemporary science, hyperauthorship is rife (Cronin, 2005) and it is rare for a single scientist to be responsible for a major theoretical breakthrough (Wuchty, Jones, & Uzzi, 2007). In the post-WWII era, big science and the large amounts of money required for research have fostered an environment that encourages research collaboration and the accordant marginalization of the solitary "genius" (Simonton, 2010; Wuchty, Jones, & Uzzi, 2007). Most research leading to Nobel Prizes is also the result of collaboration (Zuckerman, 1967), despite the anachronistic fact that Nobel Prizes cannot be attributed to more than three individuals.

The demise of the single-authored paper in scholarly communication had long been predicted (Price, 1963), and in the hard sciences, there is evidence that "much of the cutting-edge work these days tends to emerge from large, well-funded collaborative teams involving many contributors" due to the increasing specialization witnessed in all research fields (Simonton, 2013, p. 602). In parallel with the rise in the number of authors, we have also observed a growth in the number of internationally co-authored papers (Larivière, Gingras & Archambault, 2006; Sonnenwald, 2007), and many studies have shown a correlation between collaboration and impact at the micro, meso, and macro levels (Franceschet & Costantini, 2010; Narin, Stevens, & Whitlow, 1991). For example, Wuchty, Jones, and Uzzi (2007) found that while "solo authors did produce the papers of singular distinction…in the 1950s…the mantle of extraordinarily cited work has passed to teams by 2000" (p. 1038).

Although a variety of studies have demonstrated a connection between scientific impact and the various types of collaboration, no study has yet looked at these relationships from a historical standpoint. The aim of this paper is to perform an analysis of relationship between collaboration and impact using a dataset of 32.5 million papers and 515 million citations received over the 1900-2011 period. With the aid of this historical dataset, we provide empirical data on the evolution of the various types of collaboration since 1900 and perform the first historical analysis of the effect of these various forms of collaboration on citation rates, assessing as well the role of self-citations. The objective is to understand, quantitatively, whether the relationship between collaboration and impact has been static across the century. Such stability would suggest a structural relationship between these two variables, that was unaffected by the rise of citation indices or the fervor of research assessment exercises in the late 20$^{th}$ century. Furthermore, analysis of the relationship between impact and collaboration is necessary to make evidenced-based decisions about the allocation of funding and other resources for team science.

Following standard practice, we use co-authorship (that is, the presence of more than one author on the byline of a scientific publication) as our operationalization of the concept of scientific collaboration. However, the specific terminology has been disputed. Laudel (2002) argued that using co-authorship as a proxy for collaboration is based on the faulty assumption that all co-authors are also collaborators and, conversely, that all those who collaborated were listed as co-authors. Katz and Martin (1997) suggested that "[w]hat constitutes a collaboration therefore varies across institutions, fields, sectors and countries, and very probably changes over time as well" (p. 16). However, despite these caveats, "co-authorship in publications is widely considered as a reliable proxy for scientific collaboration" (Franceschet & Costantini, 2010, p. 541) and will be employed here.

**Background**
The trend of increasing co-authorship is not a new one nor is the study of the role of collaboration in science (e.g., Hagstrom, 1965). As early as 1963, Price predicted that scholarly publications will "move steadily toward an infinity of authors per paper" (p. 89). Recent studies have confirmed that co-authorship is becoming increasingly common across all disciplines (Cronin, Shaw, & Barre, 2003; Francheschet & Costanini, 2010; Galison, 2003; Larivière, Gingras, & Archambault, 2006; Persson, Glänzel, & Danell, 2004; Wuchty, Jones, & Uzzi, 2007). There is also an upwards trend in the *number* of authors credited on a paper, which sometimes reaches triple digits (Abramo, D'Angelo, & Di Costa, 2009).

*The collaboration advantage*
Several scholars have noted that collaboration is well-suited to the increasingly narrow focus scientific research (Franceschet & Costantini, 2010; Simonton, 2013), although it has been argued that increasing specialization cannot fully account for the growth in collaboration (de B. Beaver & Rosen, 1978). Other scholars have postulated that collaboration is the only practical solution when one considers the shortage of necessary resources (Wray, 2002), and it has been suggested that "easier access to public financing; aspirations for greater prestige and visibility resulting from collaboration with renowned research groups; and opportunities to attain higher productivity" are other factors that encourage collaboration (Abramo, D'Angelo, & Di Costa, 2009, p. 156). This is, however, not without its complications. For example, it has been suggested that there may be long-term negative effects when nations engage excessively in collaborations, rather than constructing their own research capabilities (Wagner & Leydesdorff, 2005).

Price and de B. Beaver (1966) somewhat humorously suggested that the social function of collaboration is to provide "a method for squeezing papers out of the rather large population of people who have less than a whole paper in them" (p. 1015). Essentially, collaboration allows for people with different (and ideally complementary) skills to come together in order to solve a single problem (Franceschet & Costanini, 2010). This integration, of course, is not always seamless and can cause friction, wasted time, and possible threats to the quality of the work when understanding is not reached by all participants (Franceschet & Costantini, 2010). Collaboration also relies on a healthy balance of trust and bureaucracy (Shrum, Genuth, & Chompalov, 2007)—the "micropolitics of collaboration" (Atkinson, Batchelor, & Parsons, 1998, p. 260) that must be negotiated for productive collaboration. Furthermore, collaboration complicates notions of contribution and responsibility in publication (Birnholtz, 2006; Kennedy, 2003).

Collaboration has been positively correlated with many metrics of academic quality (see Sugimoto, 2011 for a review). For example, collaboration has been shown to lead to higher productivity (Abramo, D'Angelo, & Di Costa, 2009; Bordons, Gomez, Fernandez, Zulueta, & Mendez, 1996; Landry, Traore, & Godin, 1996; Mairesse & Turner, 2005), with productivity increasing as team size increases (Adams, Black, Clemmons, Paula, & Stephan, 2005). The citation advantage of multi-authored papers is another incentive for scholars: many studies have demonstrated that co-authored papers tend to have higher citation impact than single-authored papers (e.g., Wuchty, Jones & Uzzi, 2007). Similarly, collaborations between industries and universities (Lebeau, Laframboise, Larivière, & Gingras, 2008) and international collaborations (Franceschet & Costantini, 2010; Glänzel, 2001; Katz & Hicks, 1997) have also been shown to yield, on average, higher scientific impact.

*The geography of collaboration*
Despite the "falling cost and growing ease of communication" among scientists (Katz & Martin, 1997, p. 8), there is evidence of a "'proximity effect,' whereby collaboration intensity is inversely proportional to the distance between the players at stake" (Abramo, D'Angelo, & Di Costa, 2009, p. 156; see also Cronin, 2008; Gieryn, 2002; Katz & Martin, 1997; Sugimoto & Cronin, 2012; Yan & Sugimoto, 2011). To incentivize scholars to collaborate across geographic boundaries, a number of institutional and governmental initiatives have been put into place (Abramo, D'Angelo, & Di Costa, 2009).

Scholars have the potential to gain academic capital for engaging in collaboration, and a number of studies have demonstrated a citation advantage for articles co-authored across institutions and nations (see Ganzi, Sugimoto, & Didegah [2012] for a review of this work). However, this advantage is not universal. Frame and Carpenter (1979) suggested that international collaboration is more likely to be witnessed in "basic" fields, and that "extra-scientific factors (for example, geography, politics, language) play a strong role in determining who collaborates with whom in the international scientific community" (p. 481). Data on the proportion of papers written in international collaboration also shows that this proportion is lowest in fields that have more local values, like social sciences, engineering, clinical research, and highest in disciplines who are more universal in their objects like mathematics, physics and space science (Gingras, 2002). Variation is also seen at the country level, where countries with weaker scientific infrastructure tend to engage more heavily in international collaboration (Luukkonen, 1992). Another (perhaps more intuitive) finding was that "the larger the national scientific enterprise, the smaller the proportion of international co-authorship" (Frame and Carpenter 1979, p. 481). There is compelling evidence that the geographic proximity between the first and last author generate higher citations (Lee, Brownstein, Mills, & Kohane, 2010) and that international collaborations, in general, generate higher citations (Glanzel, 2001).

*Self-citations*
Self-citations have been critically viewed as a gaming mechanism in scholarly communication (MacRoberts & MacRoberts, 1989), and several studies have examined the prevalence of self-citations at multiple levels of analysis, including papers, journals, individuals, and countries (Eto, 2003; Frandsen, 2007; Minasny, Hartemink, & McBratney, 2010; Snyder & Bonzi, 1998; Tagliacozzo, 1977). Early studies found that self-citations ranged from 8% at the individual level to 20% at the journal level (Garfield & Sher, 1963), while more recent studies have given percentages as high as 36% (Aksnes, 2003). In addition, some studies have demonstrated that the rate of self-citation fluctuates between disciplines (e.g., Bonzi & Snyder, 1990). Although complaints have been leveled at self-citation practices, Glänzel, Debackere, Thijs, and Schubert (2006) found that, at the macro-level, "there is no reason for condemning self-citations in general or for removing them from citation statistics" (p. 275).

The citation advantage of co-authored works has been challenged on the grounds that it simply results from the "amplification" of the known practice of self-citation (van Raan, 1998): that is, if each author self-cites to the same degree that a single-author would, the citations to a co-authored paper should be multiplied by the number of co-authors. It should be no surprise that self-citations increase with the number of co-authors (Wallace, Larivière, & Gingras, 2012), although it must be noted that this citation rate does not increase linearly (Glanzel & Thijs, 2004), suggesting that self-citation is not a sufficient explanation for the citation advantage of co-authored works. Instead, a specific citation impact seems to be associated with collaborative work (van Raan, 1998).

Just as collaboration practices vary by discipline (Larivière, Gingras & Archambault, 2006), so too does the citation impact of collaboratively written works (Abramo, D'Angelo, & Di Costa, 2009; Pečlin, Južnič, Blagus, Sajko, & Stare, 2012). These studies often lack generalizability due to small sample sizes, disciplinary focus, and limited time periods for analysis. Additional large-scale research is needed to identity the extent to which a citation advantage prevails across time and discipline. In addition, despite evidence of differences in self-citation by subspecialties, age, and (to a lesser extent) gender (Hutson, 2006), there is still much research to be done on the interaction between self-citation and impact.

**Methods**
The data for this paper are drawn from Thomson Scientific's Science Citation Index Expanded (SCIE), Social Sciences Citation Index (SSCI), and Arts and Humanities Citation Index (AHCI) for the 1900-2011 period. We analyze 28,160,453 papers (articles, notes and reviews) and 484,393,178 citations received in Natural and Medical Sciences (NMS) as well as 4,347,229 papers and 30,587,347 citations received in the Social Sciences and Humanities (SSH). The evolution of the relationship between scientific impact and three types of collaborations is also presented. These three types of collaboration are: 1) co-authorship (i.e., number of authors), 2) interinstitutional collaboration (i.e., number of addresses), and 3) international collaboration (i.e., number of countries). While co-authorship data are presented here from 1900 onwards, interinstitutional and international collaboration data are only available from 1973 onwards, as it was in that year that Thomson Reuters' predecessor -- the *Institute for Scientific Information* -- began to index institutional addresses in a consistent manner.

Citations are counted from publication year until the end of 2011. In order to have a citation window of at least two years following the initial publication year, scientific impact data are presented up until 2009. To take into account different citation practices across subfields of science and publication year, all citation data were normalized according to the average number of citations received by the papers that were published in the same year and in the same speciality (average of relative citations – ARC)[1]. Accordingly, we have ensured that the "collaboration" variable is isolated and that the greater impact of collaborative research is not due to disciplines with higher collaboration rates having greater citation traffic. In order to assess whether self-citations played a role in the (greater) impact of collaborative research, two types of field-normalized citation impact were compiled, one including self-citations and the other excluding self-citations. In the latter case, all authors' self-citations – irrespective of their order in authors' list – were excluded from each paper in the numerator as well as from the denominator itself (i.e., the average number of citations of all papers in the respective specialty that were published in the same year). We did *not* remove self-citations at the institutional or country level when analyzing the effect of the evolution of the number of addresses or countries on scientific impact, as it is *individuals* who cite, not institutions or countries. We also compiled data (not shown) on the top 5% most cited paper for each of the specialties—the trends were identical to those shown in the figures.

**Results**
*Evolution of collaboration*
For all types of collaboration analysed, the numbers of authors, addresses, and countries were grouped into classes. Figure 1 presents the yearly evolution of the percentage of papers in each of the classes of number of authors, number of institutional addresses, and number of countries. It shows that, in all collaboration types, single author/address/country papers are decreasing, both in NMS and SSH, a finding that has been shown in other studies (Larivière, Gingras & Archambault, 2006; Wuchty, Jones, & Uzzi, 2007). More specifically, papers with one author accounted, in 1900, for 87% and 97% of all papers in NMS and SSH respectively; these percentage are, in 2011, 7% and 38% respectively. At the beginning of the period, the decrease of the proportion of single-author papers in NMS is due to the increase of papers with two authors; the proportion of the latter has also decreased since the beginning of the 1960s, mainly due to the increase of papers with more than two authors. The same phenomenon is observed for papers with 3 authors at the beginning of the 1980s. In 2011, paper classes with 4-5 and 6-10 authors account for about 29% and 27% of all NMS papers, respectively. Classes with 11-20 authors increased their proportion of papers by more than 2000% (0.22% to 4.5%) between 1980 and 2011, while papers with 21 authors or more increased by more than 1000% (0.04% to 0.5%) over the same period. In the SSH, all paper classes with more than one author increase, although, in 2011, the *mode* (i.e., the most common number of authors) is still one.

---
[1] In order to have robust trends in the graphs, only ARC scores based on at least 100 papers are shown.

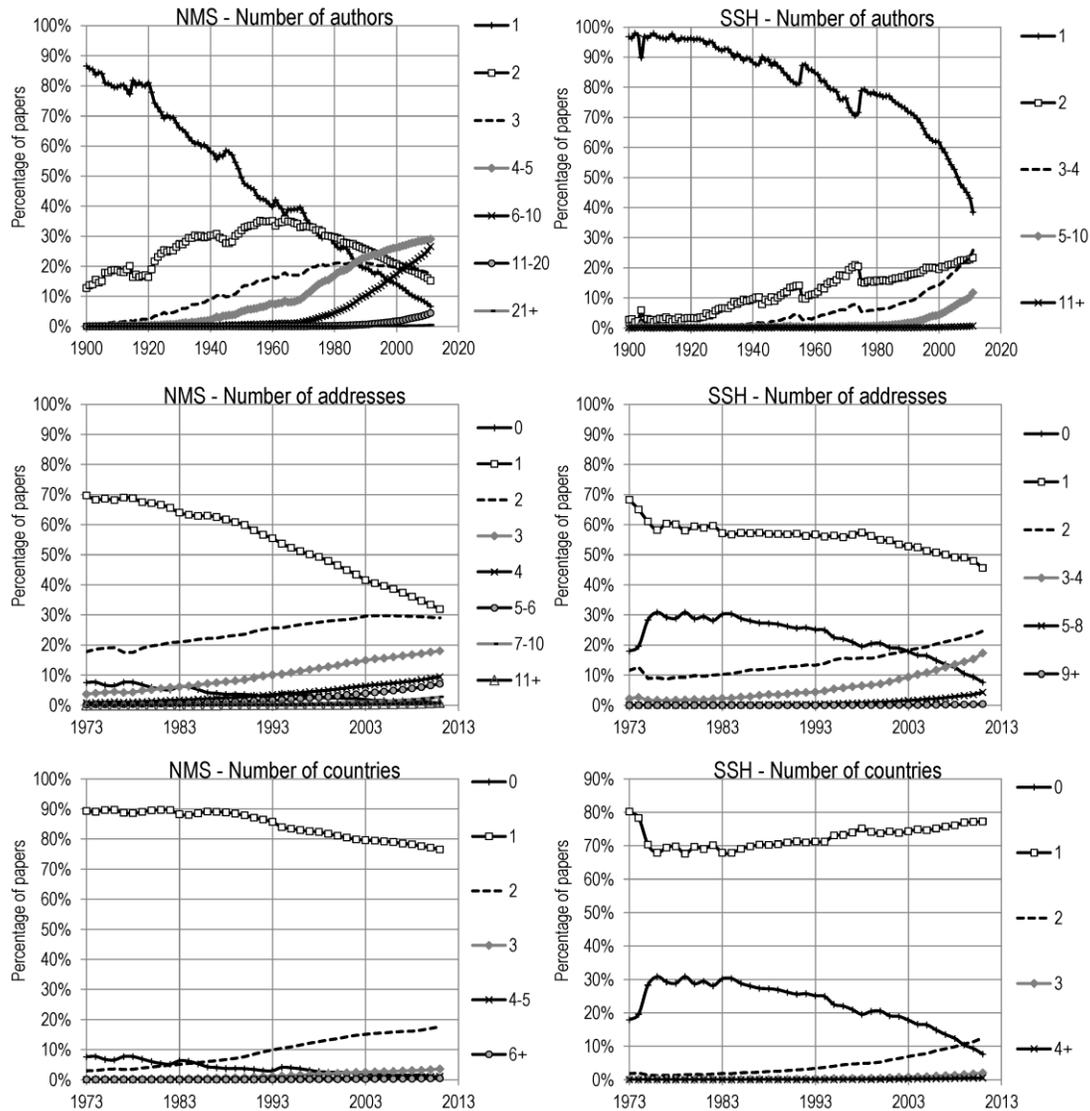

Figure 1. Percentage of papers by classes of numbers of authors, addresses, and countries, for natural and medical sciences (NMS) and social sciences and humanities (SSH), 1900-2011

We also observe a decline of the proportion of papers with only one address, from 70% in 1973 to slightly more than 30% in 2011 in NMS and from 70% to 46% in SSH over the same period of time. In NMS, we observe a stabilization of the share of papers with two addresses while papers with three addresses or more are still increasing. Given that SSH journals' editorial policies sometimes do not specify the institutional address of authors, approximately 30% of papers from 1975 lacked institutional addressed. This percentage has decreased steadily, although it still accounts for 8%. In NMS, papers without any address account for only 1% of all papers, down from 8% in 1973. Some of these papers are actually indexing mistakes, and generally happen in lower-tier journals.

Papers with authors from two countries also increased their proportion of all published literature, accounting for 18% of NMS and 14% of SSH literature in 2011. In NMS, papers with three, four-five, and six or more authors have increased by 2651%, 4969%, and 8365% between 1973 and 2011 while, in SSH, papers with three and four authors or more increased their share of all papers by more than 3000% and 4300%, respectively. Single country papers are still the mode, accounting for 77% and 84% of all papers in NMS and SSH, respectively. These data show an increase of all types of collaboration, both in NMS and SSH. The following sections will assess how these different types of collaboration and their intensity have influenced papers' scientific impact over the course of the last decades.

*Inclusion vs. exclusion of self-citations*
It is commonly thought that the larger impact of collaborative research is due, at least in part, to authors' self-citations (Herbertz, 1995). If a paper contains 20 authors, and each of the authors cites it at least once, then it accumulates 20 self-citations. At the other end of the spectrum, a paper with only one author, who cites the paper in a following publication, will result in only one self-citation. Data presented in Figure 2 show the gain in scientific impact (field-normalized citation rates) obtained by including self-citations as a function of the number of authors. More specifically, Figure 2 shows that, in both areas, there is a loss in citation impact when self-citations are included for non-collaborative research (i.e., one author). In SSH, papers with two authors obtain similar normalized citation rates whether or not self-citations are included, whereas papers with at least three authors enjoy a steady increase in impact, gained from including self-citations. This gain reaches 20% at 25 authors, and oscillates around that percentage until 40 authors appear on the paper. In NMS, it takes many more authors in order to benefit from self-citations. Papers with fewer than four authors obtain, on average, lower field-normalized scores when self-citations are included than when they are excluded, which is due to the fact that the self-citations are excluded from both the numerator and the denominator. It is only when a paper has at least five authors that ARC scores including self-citations rise above those without self-citations. The gain in impact from self-citations is much smaller in NSE than in SSH for the same number of authors, which is likely a consequence of the lower number of citations in these disciplines. These results are consistent with those obtained by Aksnes (2003) in an analysis of Norwegian papers. In order to remove this self-citation effect—even though it is small—the data presented in the rest of the paper exclude all authors' self-citations.

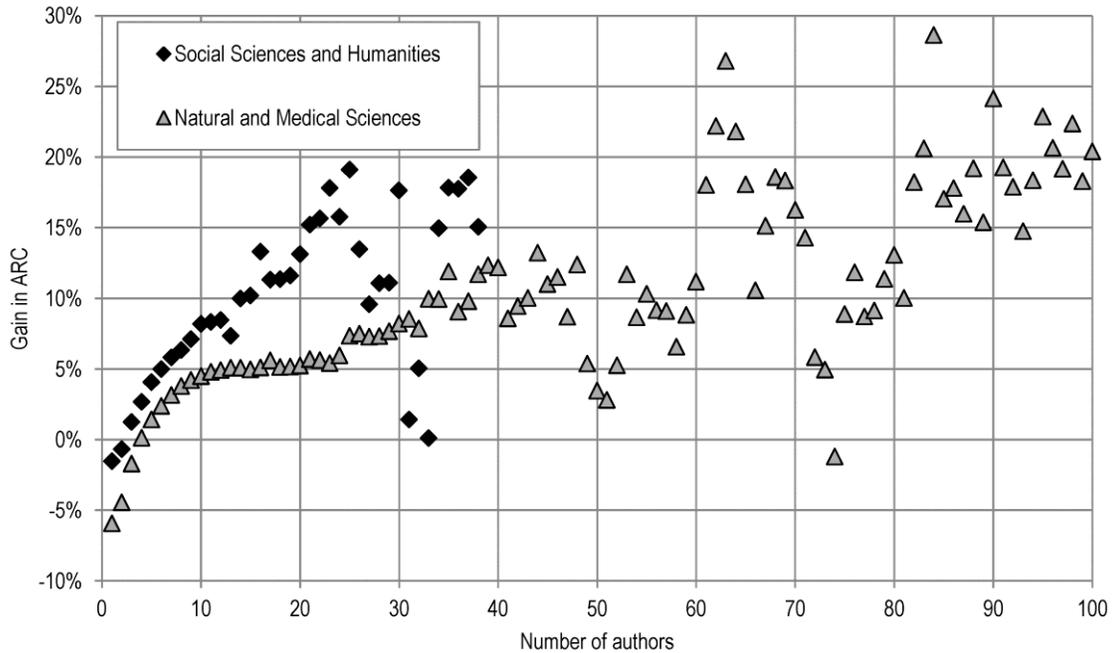

Figure 2. Gain in average of relative citations (ARC) when self-citations are included, as a function of the number of authors in NMS and SSH, 2005-2009. Three-year moving averages.

***The relationship between scientific impact and the number of authors, addresses, and countries***

Figure 3 presents the field-normalized impact of papers for both NMS and SSH, excluding self-citations, as a function of the number of authors, addresses, and countries. In NMS, the impact score of papers increases steadily with the number of authors until it reaches about 45 authors, where it flattens and oscillates, although it remains well above average. The same phenomenon is also observed in the social sciences. Given the lower proportion of collaborative research papers in those disciplines, as well as the lower number of papers involved, the impact score tends to oscillate at the level of 20 authors – although, again, it remains above average. Compared to SSH, many more authors are required in NMS in order to realize a given percentage of citation gain from collaboration. SSH start gaining citations with three authors with a roughly linear gain until 20 authors, whereas NMS papers gain citations starting with 5-author papers with no additional gain in the 10-20 author range. Similar trends are observed when considering the numbers of addresses and countries: the larger number of addresses and countries appearing on a paper the larger the impact. Unsurprisingly, papers with no address obtain the lowest impact.

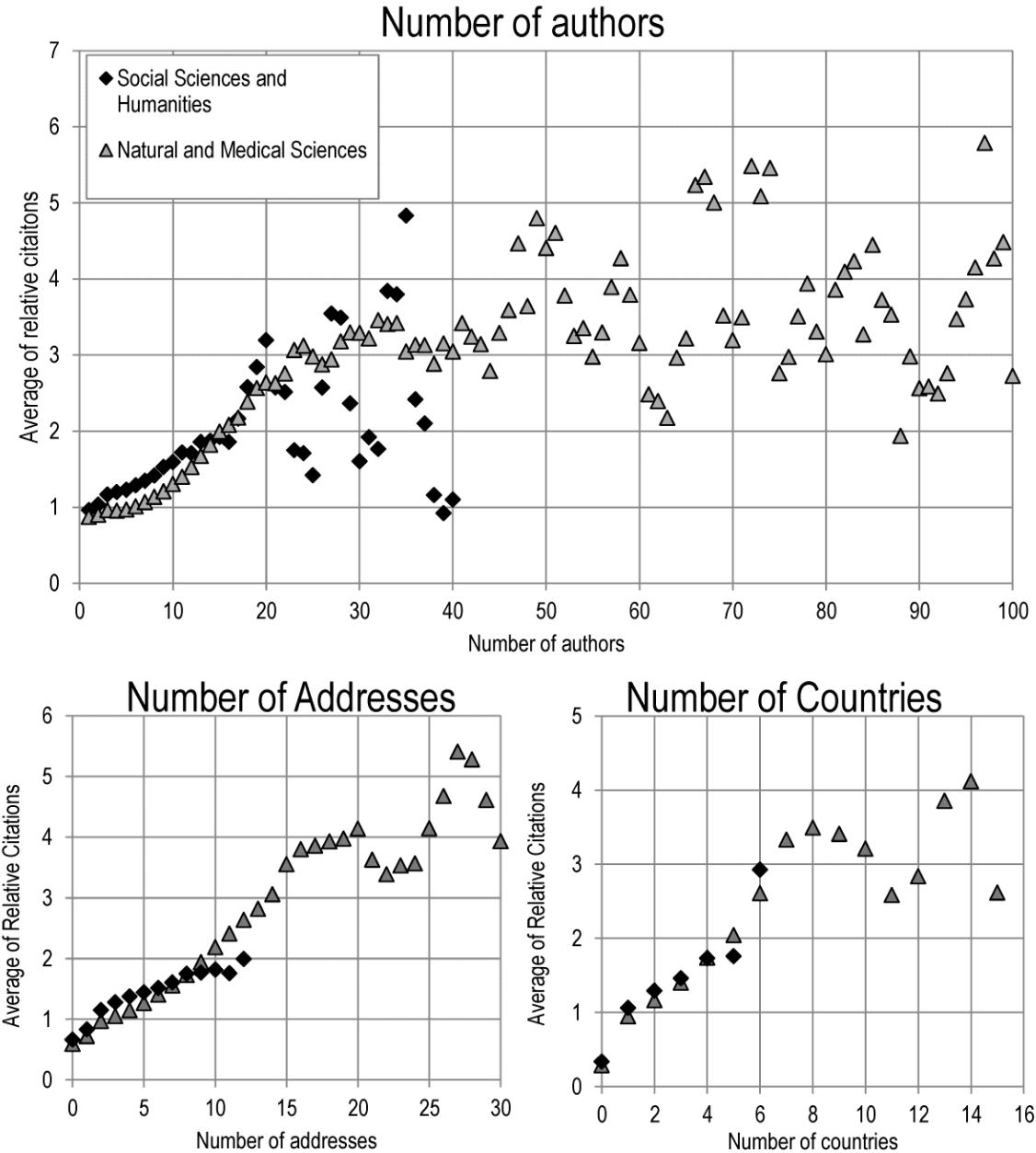

Figure 3. Average of relative citations (ARC) with self-citations excluded, as a function of the number of authors, addresses, and countries in NMS and SSH, 2005-2009. Three-year moving averages. Only ARC scores based on 100 papers or more are shown.

*Historical evolution of the impact of collaborative research*
Figure 4 presents ARC scores for the 1900-2009 period, excluding self-citations, of NMS and SSH papers for specific classes of author counts. In both areas and for all publication years, we observe that, on average, papers with fewer authors consistently obtain lower citation rates. It also shows that there is an inflation of the average number of authors associated with a higher scientific impact. For example, although papers in the NMS with two or three authors have an impact larger than the world average in 1900, this has no longer been the case since the first decade of the 2000s. Since the early 1980s, papers with fewer than 21 authors experience a decrease in their mean citation rates. A similar trend is also observed for SSH, although a lower number of authors is needed, on average, to obtain a larger impact. We also see that a larger number of authors yields a larger comparative impact in the NMS than in SSH.

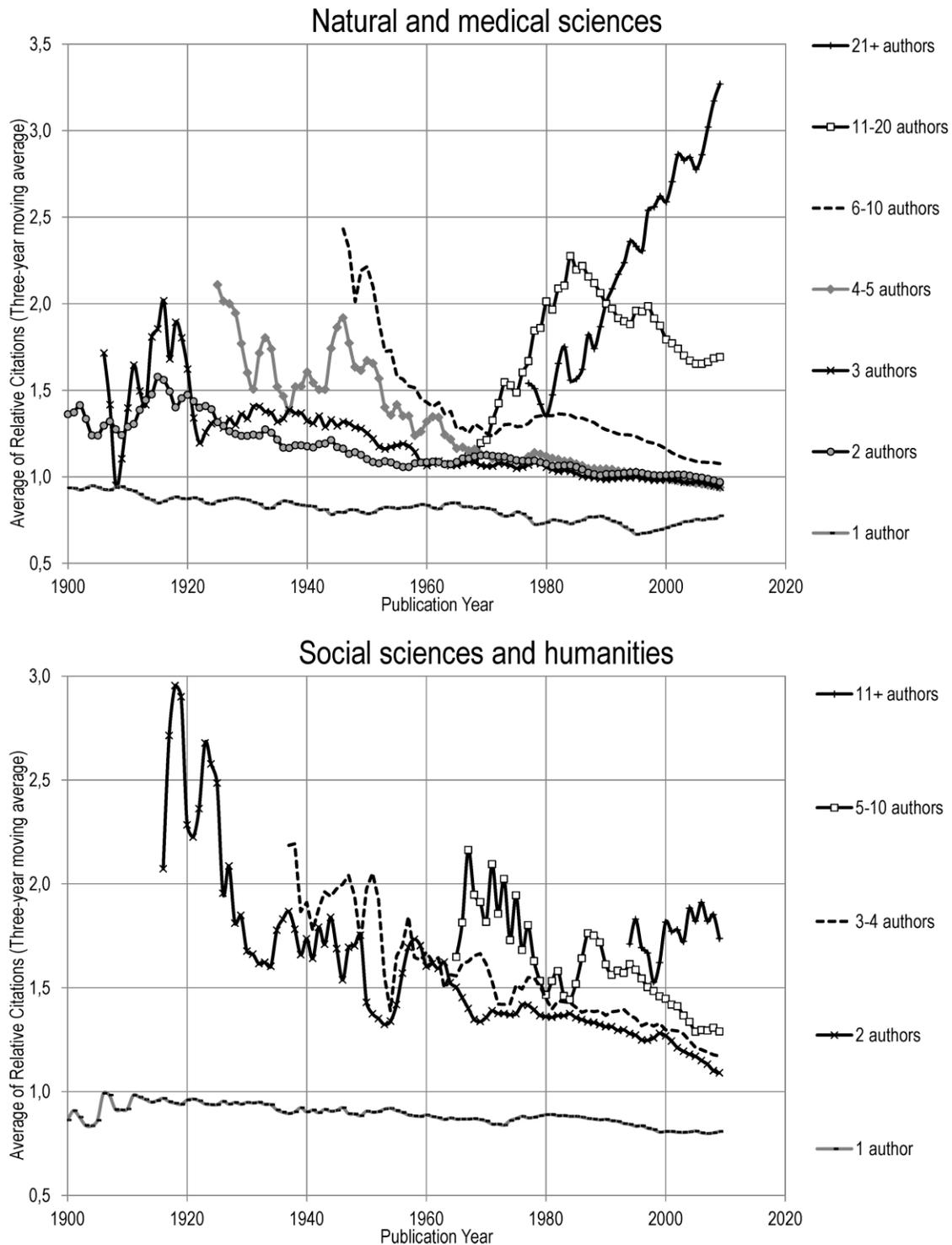

Figure 4. Average of relative citations (ARC) of papers in Natural and Medical Sciences and

Social Sciences and Humanities, as a function of their numbers of authors, 1900-2009. Three-year moving averages. Only ARC based on 100 papers or more are shown.

Figures 5 and 6 present the relationship between the number of addresses (Figure 5), countries (Figure 6), and scientific impact for the 1973-2009 period. It shows that impact increases with the number of addresses or countries, and that this relationship is consistently observed both over time and between research areas. We also observe an inflation in the number of addresses associated with higher scientific impact: over the 1973-2009 period, a decrease in impact is witnessed for papers with fewer than 11 addresses (in NMS) and papers with fewer than nine addresses (in SSH). Similar trends are also observed when the numbers of countries is taken into consideration. On both figures, papers without any address obtain a scientific impact below average throughout the period.

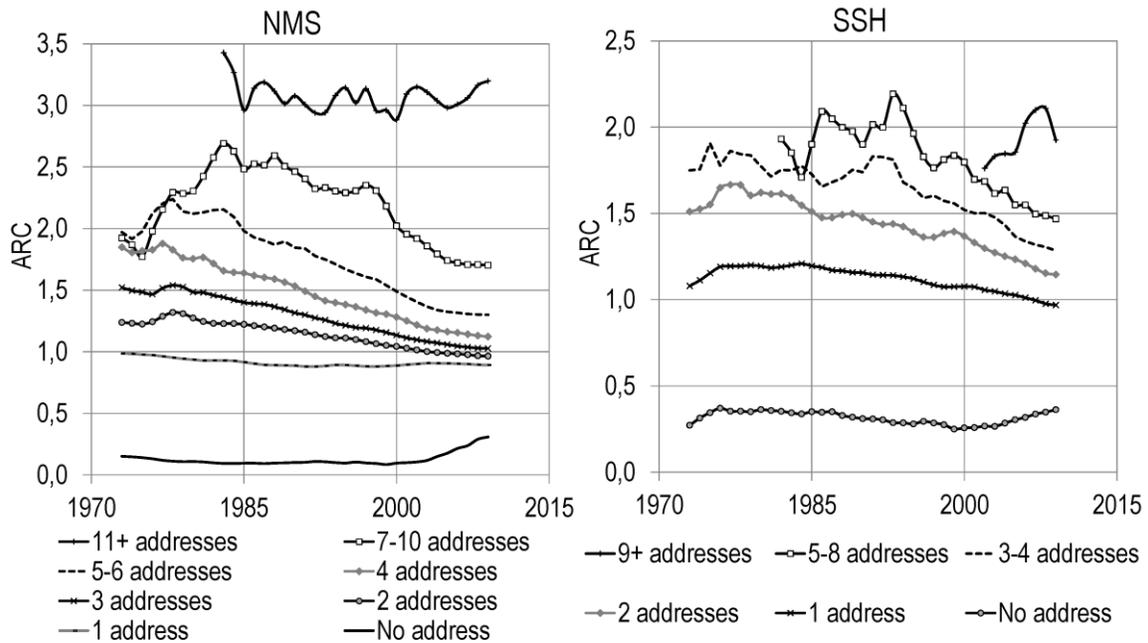

Figure 5. Average of relative citations (ARC) of papers in the natural and medical sciences and social sciences and humanities, as a function of their numbers of addresses, 1973-2009. Three-year moving averages. Only ARC based on 100 papers or more are shown.

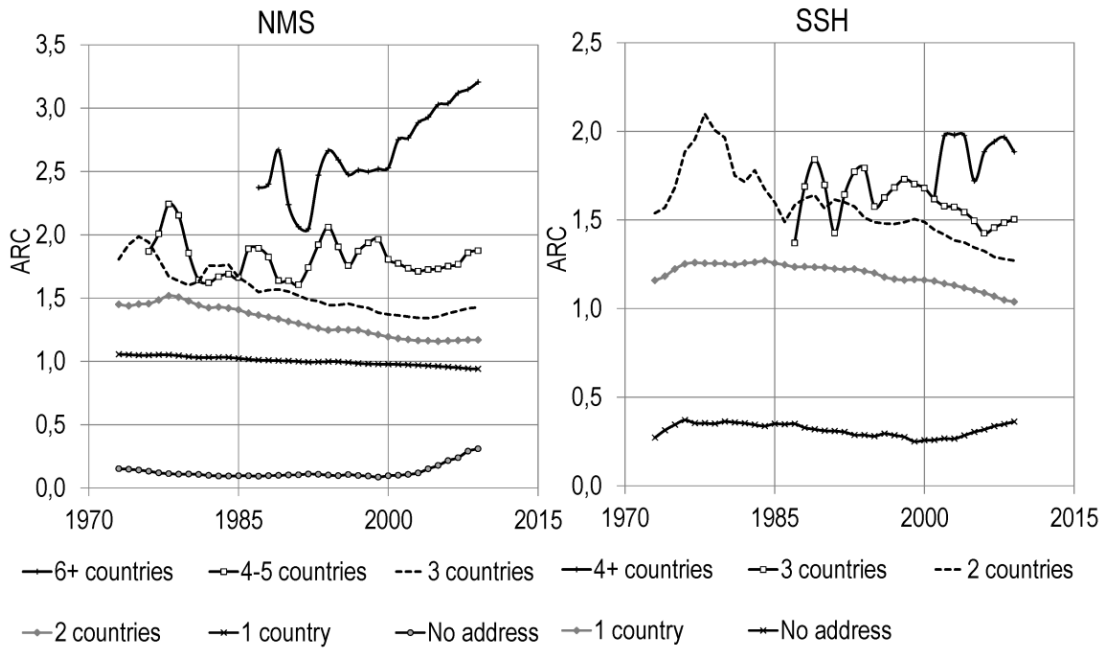

Figure 6. Average of relative citations (ARC) of papers in the natural and medical sciences and social sciences and humanities, as a function of their numbers of countries, 1973-2009. Three-year moving averages. Only ARC based on 100 papers or more are shown.

**Discussion and conclusion**
The results presented above show that, from 1900 onwards, co-authorship, inter-institutional collaboration, and international collaboration have been increasing in both NMS and SSH. More specifically, single-authored papers decreased in NMS from 87% in 1900 to 7% in 2011 and, in SSH, from 97% to 38% over the same period. At the beginning of the period in question, the decrease of single-authored papers in NMS is due to the increase of papers with two authors, the proportion of the latter has also decreased since the beginning of the 1960s, mainly in favor of papers with more than two authors. Papers with one address have also been decreasing, accounting in 2011 for 32% and 46% of all papers in NMS and SSH, respectively. Hence, for both domains, the majority of contemporary papers are the result of inter-institutional collaboration. However, despite its increase throughout the period, international collaboration remains, at the global level, a relatively marginal phenomenon: in 2011, 22.7% and 16.4% of all papers in NMS and SSH, respectively. Similarly, multilateral collaboration—that is, collaborations involving more than two countries—only accounted, in 2011, for 5.1% of all papers in NMS and 2.8% of all papers in SSH. Our findings suggest that while collaboration is becoming ubiquitous across fields, international collaboration has not seen the same gains as inter-institutional collaboration. This may be a result of personal mobility (particularly for women [see Larivière, et al., 2013]) and the national funding policies of large countries such as the U.S.

Our results also provide evidence that the gain in self-citations that papers can obtain increases as a function of the number of authors, but flattens once a certain number of authors is reached (10-12 in NMS and 25 in SSH). In addition, this gain is higher in SSH than in NMS. Finally, the scientific impact of papers, excluding self-citations, increases with the number of authors, addresses, and countries appearing on a paper (at least for the 2005-2009 period). This suggests that self-citation contributes to, but does not fully explain, the relationship between impact and collaboration.

By providing data on the relationship between co-authorship and scientific impact for the first 50 years of the 20$^{th}$ century, we expand on the work done by Wuchty, Jones, and Uzzi (2007) and show that, as early as 1900, co-authored papers were more heavily cited than sole-authored papers. However, with the inflation in the number of contributors over the course of the century, an increasingly high number of authors is needed in order to obtain a given level of citations. For instance, while in 1919, papers with three authors had an average impact that was twice the world average, in 1950, the same impact was obtained by papers that had between 6-10 authors. In 2009, only papers with more than 21 authors reached an average impact that was twice the world average. For this group of papers, the average impact in 2009 was actually more than three times the world average. The same "inflation" phenomenon is also observed post-1973, when one looks at the number of addresses and the number of countries. In other words, when it comes to the relationship between collaboration and scientific impact, size matters. However, the top tier of papers with the highest number of addresses and countries did not gain as much in terms of citations as did the top tier in terms of the numbers of authors.

These results show that, overall, collaborative research results in higher citations rates. This may also suggest that this relationship is not a "mechanical" artifact caused by an increase in self-citations, but rather an effect of the greater epistemic value associated with collaborative research (Wray, 2002). Although there is no single reason that can explain this relationship, one popular hypothesis is that the most important scientific problems are complex and can only be solved by a team of researchers having complementary expertise (de B. Beaver, 2004; Wray, 2002). For example, the pooling of different countries' human and financial resources in the creation and use of particle accelerators can actually produce higher scientific impact. Along the same lines, the equipment needed to carry out such work cannot be handled by a single researcher, let alone by a single researcher in one area of specialization. One could also mention work in genomics which involve many institutions and countries. In addition, the complexity of scientific problems often leads to the forming of interdisciplinary research teams which generate results that are in turn applicable to each of the areas involved. Finally, the intersubjectivity associated with collaboration ensures that the resultant knowledge reflects a greater level of consensus and potentially greater epistemic value than research conducted by an isolated researcher. Future research should seek to test these hypotheses and further investigate what attributes of collaborative research contribute to the increased impact of the scientific work it produces.